\documentstyle[12pt]{article}
\begin{document}
\baselineskip=0.6 cm
\newpage
$\,$

{\bf Wavelet Analysis of Inhomogeneous Data with application to
 the Cosmic Velocity Field}

\vspace {0.5 cm}

S. Rauzy$^1$, M. Lachi\`eze-Rey$^2$ and R.N. Henriksen$^3$

$^1$ Universit\'e de Provence and Centre de Physique Th\'eorique,
C.N.R.S. Luminy, Case 907, F-13288 Marseille Cedex 9, France
\\\indent $^2$ Service d'Astrophysique, C.E.N. Saclay, F-91191 Gif-sur-Yvette
Cedex, France
\\\indent $^3$ Astronomy Group, Dept of Physics, Queen's University,
Kingston, Ontario K7L 3N6 Canada

\vspace {0.5 cm}

\noindent {\bf Abstract.}
In this article we give an account of a method of smoothing spatial
inhomogeneous data
sets by using wavelet reconstruction on a regular grid in an auxilliary space
onto which the original data is mapped. In a previous paper by the present
authors, we devised a method for inferring the velocity potential from the
  radial component of the cosmic velocity field assuming an ideal sampling.
  Unfortunately the sparseness of the real data as well as errors of
measurement
  require us to first smooth the  velocity field as observed on a 3-dimensional
  support (i.e. the galaxy positions) inhomogeneously distributed throughout
  the sampled volume. The wavelet formalism permits us to introduce a minimal
  smoothing procedure that is characterized by the variation in size of the
  smothing window function. Moreover the output smoothed radial velocity field
  can be shown to correspond to a well defined theoretical quantity as long as
  the spatial sampling support satisfies certain criteria. We argue also that
  one should be very cautious when comparing the velocity potential derived
  from such a smoothed radial component of the velocity field with related
  quantities derived from other studies (e.g : of the density field).

\newpage

$\,$
\vspace {0.3 cm}

\noindent {\bf 1. Introduction}
\vspace {0.3 cm}

An `inverse' problem posed frequently in cosmology is to construct
the velocity potential (assuming an irrotational velocity field) from
observations of only the `radial' component of the velocity field. We have
argued elsewhere [7] that wavelet
transforms provide a convenient formalism for this procedure, but subsequently
in the course
of our practical investigations we found that the signal is easily
distorted by that due to the inhomogeneity of a non-uniformly sampled data set.
Such a
sample is constituted by a set of galaxy peculiar velocities
distributed (as they in practice always are) over a nearly random support, and
it
is to this example that we address most of our remarks in this paper. However
the solution that we propose, namely a non-linear algorithmic mapping from
the real space into a fictitious but uniformly sampled space may be of general
interest and is therefore reported here.

The ultimate cosmological objective is
to extract information concerning the fluctuations of the total
mass in the universe $\delta_t({\bf x)}$ from the observed radial component
$v_r({\bf x})$ of the cosmic
peculiar velocity field ${\bf v}({\bf x})$, assuming that the latter has been
created
by fluctuations in the self-gravitating matter. One of the goals is indeed,
by comparing the fluctuations of the total mass obtained as above
with the fluctuations
of the luminous mass $\delta_l({\bf x})$ obtained from the study of the
distribution of the galaxies in the universe, to characterize the
distribution of the dark matter in the universe $\delta_d({\bf x})\,
=\,\delta_t({\bf x})\,-\,\delta_l({\bf x})$ .

According to the standard models of large scale structure formation,
there are good reasons to believe that the cosmic velocity field
is irrotational at least above some ill-defined scale $s_c$.
Bertschinger and Dekel [1] have pointed out that consequently above the scale
$s_c$, the cosmic
velocity field ${\bf v}({\bf x})$ may be derived  from a potential
(${\bf v}({\bf x})\,=\,\nabla\,{\Phi}({\bf x})$, i.e : ${\bf v}$
is curl-free), and that this kinematical potential ${\Phi}$ can
be extracted by integrating the  observed radial
component of the velocity field $v_r({\bf x})$ along the line-of-sight :
$${\Phi}({\bf x})\,\,=\,\,(Pv_r)({\bf x})\,\,
=\,\,{\int_0^1\,dl\,v_r(l\,{\bf x})} \eqno(1.1)$$
Moreover, above the scale $s_c$, the velocity fluctuations still belong to the
linear regime and
the total mass density field $\delta_t({\bf x})$ is thus linked
to the kinematical potential $\Phi({\bf x})$ through the Poisson
equation ($\delta_t({\bf x})\,\,\propto\,\,\nabla^2\Phi({\bf x})$).

In our previous paper [7],
we devised a method based on the properties of the wavelet transform
for inferring the kinematical potential from the observed radial
component of the cosmic velocity field. We summarize in section 2
why our method is well adapted for solving the problem of the choice of
the  scale $s_c$ ,and how we can directly derive from the
radial velocity field quantities such as the laplacian of the
kinematical potential (and thus the fluctuation in the total mass).

Unfortunately, as was observed above, the radial velocity field is measured
with large statistical
uncertainties and is sampled on a support (i.e : the positions of
galaxies) inhomogeneously distributed throughout space.
Thus one has to first smooth the observed radial velocity field
before applying no matter which inversion procedure to find the
kinematic potential and hence the reconstructed complete velocity field.
The POTENT inversion [2,4]
has already given impressive
results. However, we see two limitations in the
smoothing scheme used in the POTENT procedure. Firstly, the size
of their smoothing window function doesn't vary spatially,
and so fictitious information is added in undersampled regions and
a significant part of the signal
is lost in oversampled regions. Secondly, the errors
intervening during the smoothing procedure of the POTENT method
are difficult to quantify. In this article, in section 3, we present
a smoothing
procedure based on the wavelet analysis, which permits one to {\it optimally}
smooth
a field sampled on a 3-dimensional
support distributed inhomogeneously throughout space.
The natural variation in size of the smoothing window function in our procedure
permits
us to realize a minimal or optimal smoothing procedure
(i.e : without loss of information).
Moreover we prove that our smoothed output field corresponds to
a theoretical quantity (defined by the wavelet analysis formalism), so
permitting us to quantify and control the errors intervening during
the successive
steps of our smoothing procedure.

Finally, we analyze in a general way in section 4
the wavelet transform procedure involved in
the reconstruction of the kinematical potential
given the smoothed radial component of the velocity field. We show that
this operation of reconstruction doesn't commute with the preliminary
smoothing operation. This creates difficulties already at the `a priori'
theoretical level when attempting to compare
the reconstructed velocity field or its associated mass density field
with their correspondant quantities derived
from other studies, since in those cases the smoothing may be effected in a
different
order. This problem is general in the sense that it is in no way unique to our
smoothing procedure but rather depends on the global character
of equation (1.1).

\noindent {\bf 2. Reconstruction of the velocity field using wavelet analysis }
\vspace {0.3 cm}

\noindent {\it 2.1. The decomposition by scales }
\vspace {0.1 cm}

One of the characteristic features of wavelet analysis is that it allows a
simultaneous study of both the positional and the scaling properties of
a function. Thanks to the wavelet reconstruction theorem (see
[6]), a square integrable function
(for example the radial velocity field $v_r({\bf x})$ defined at each
point ${\bf x}$ of our 3-dimensional space) can be decomposed
in a family of functions $v_r^{(s)}(\bf x)$ as follows :
$$v_r({\bf x})\,\,=\,\,(Wv_r)({\bf x})\,\,=\,\,\int_{0}^\infty\,
{{ds}\over{s}}\,\,v_r^{(s)}({\bf x}) \eqno(2.1)$$
The integral is performed over all scales $s$ and
$v_r^{(s)}({\bf x})$ is the component of $v_r(\bf x)$ smoothed on the scale
$s$.
It is in fact the spatial convolution of
the radial velocity field $v_r({\bf x})$ with the "reproducing
kernel" $K(s,{\bf x},{\bf y})$ (see the illustration figure 1a),
a well-defined function
centered on ${\bf x}$ and of
spatial extension $s$ (see [7]) namely :
$$v_r^{(s)}({\bf x})\,\,=\,\int_{-\infty}^\infty \int_{-\infty}^\infty
\int_{-\infty}^\infty\,\,d^3{\bf y}
\,\,K(s,{\bf x},{\bf y})\,v_r({\bf y}) \eqno(2.2)$$
$v_r^{(s)}({\bf x})$ thus contains information localized
"around" the scale $s$ in  frequency (scale) space.
As the scale $s$ decreases, a more and more detailed picture of the velocity
field
is thus available. Because of the redundant properties of the continuous
wavelet transform, the integral involved in equation 2.1
can in practice be replaced by a discrete sum over a few scales
(see [3,7]).
Moreover, the reconstruction will necessarily be limited to a maximum
resolution
 (minimum scale $s_c$) by the finite spatial resolution of the
 data distribution.

To emphasize this latter point let us introduce two operators $W_{s_c}$ and
$W^{s_c}$ acting on $v_r$ as follows :
$$v_r\,\,=\,\,Wv_r\,\,=\,\,W_{s_c}v_r\,\,
+\,\,W^{s_c}v_r \eqno(2.3)$$
$$(W_{s_c}v_r)({\bf x})\,\,=\,\,\int_{s_c}^\infty\,
{{ds}\over{s}}\,\,v_r^{(s)}({\bf x}) \eqno(2.4)$$
$$(W^{s_c}v_r)({\bf x})\,\,=\,\,\int_0^{s_c}\,
{{ds}\over{s}}\,\,v_r^{(s)}({\bf x}) \eqno(2.5)$$
$(W^{s_c}v_r)({\bf x})$ contains the information
about $v_r({\bf x})$ at scales smaller than $s_c$ and the term
$(W_{s_c}v_r)({\bf x})$ may
be considered to be
a smoothed version of the radial velocity field determined by a smoothing
window function $g(s_c,{\bf x},{\bf y})$ of size $s_c$ namely :
$$g(s_c,{\bf x},{\bf y})\,\,=\,\,
\int_{s_c}^\infty\,\,
{{ds}\over{s}}\,\,
K(s,{\bf x},{\bf y}) \eqno(2.6)$$
\begin{figure}
\vspace{9 cm}
\caption{
For a given scale $s$ and position ${\bf x}$,
the variations on a 2-dimensional cut
of :
(a) the "reproducing kernel" $K(s,{\bf x},{\bf y})$,
(b) the "generalized kernel" $L(s,{\bf x},{\bf y})$.
}
\end{figure}
Note then how in practise one performs the reconstruction of $v_r$. The
$v_r^{(s)}(\bf x)$ components involved in the integral over the scales in
equation (2.3)
are summed step by step, proceeding from the larger cutoff scales to the
smaller ones.
When the desired level of spatial resolution is reached (for example
at a given scale $S$), we then halt the reconstruction and so obtain the
smoothed version of $v_r(\bf x)$ at this scale $S$,
$(W_{S}v_r)({\bf x})$. This procedure is particularly convenient in the
case of our study for which the radial velocity field has to be first
smoothed at an ill-defined scale $s_c$ in order to eliminate the
small-scale rotational component of the cosmic velocity field.
Our decomposition by scales permits us to add progressivly the components
of the field at smaller and smaller scales, halting the reconstruction when the
signal becomes noisy (see [7]).

\noindent {\it 2.2. Reconstructing other quantities from the radial velocity
data}
\vspace {0.1 cm}

In order to derive the kinematical potential $\Phi({\bf x})$ from the
radial velocity field $v_r(\bf x)$, we have to apply on $v_r$ the
`integral along the line-of-sight' operator $P$ (defined in equation 1.1).
We then decompose the kinematical potential by scales as done for the velocity
in equation
(2.1), and the component $\Phi^{(s)}(\bf x)$ is linked to $v_r^{(s)}(\bf x)$
by the wavelet transformed equation (1.1):
$${\Phi}^{(s)}({\bf x})\,\,=\,\,(Pv_r^{(s)})({\bf x})\,\,
=\,\,{\int_0^1\,dl\,v_r^{(s)}(l\,{\bf x})} \eqno(2.7)$$
A practical way to reconstruct the potential is now to introduce
the "generalized kernel" $L(s,\bf x,\bf y)$ (see the illustration
figure 1b) by applying the operator $P$
on the reproducing kernel $K(s,\bf x,\bf y)$ :
$$L(s,{\bf x},{\bf y})\,\,=\,\,P\circ K(s,{\bf x},{\bf y})\,\,
=\,\,{\int_0^1\,dl\,K(s,l\,{\bf x},{\bf y})} \eqno(2.8)$$
It then follows from equation (2.2)
that the component $\Phi^{(s)}(\bf x)$ is directly given by the convolution
of the observed radial velocity field $v_r(\bf x)$ by
the generalized kernel $L$ which we emphasize depends only on the properties
of the wavelet mother function and so is known in principle to arbitrary
accuracy, either analytically or numerically.
This convolution has the explicit form :
$$\Phi^{(s)}({\bf x})\,\,=\,\int_{-\infty}^\infty \int_{-\infty}^\infty
\int_{-\infty}^\infty\,\,d^3{\bf y}
\,\,L(s,{\bf x},{\bf y})\,v_r({\bf y}) \eqno(2.9)$$
The important idea is that other quantities of physical significance
may be reconstructed in
the same way, such as the tangential components of the velocity field,
the divergence of the field, the laplacian of the kinematical potential, etc
... .
For each such quantity which depends linearly on the velocity field,
a specific kernel function similar to $L$ for the potential can be introduced
that reduces the reconstruction at a given scale to a convolution with the
radial data as in equation (2.9).
For example, to the laplacian of the kinematical potential $\nabla^2\Phi$
will be associated a generalized kernel
$N(s,\bf x,\bf y)$ given by :
$$N(s,{\bf x},{\bf y})\,\,=\,\,\nabla^2_{\bf x} L(s,{\bf x},{\bf y})\,\,
=\,\,{\int_0^1\,dl\,\nabla^2_{\bf x}L(s,l\,{\bf x},{\bf y})} \eqno(2.10)$$
This feature avoids in practice the amplification of the errors arising
during the successive steps of the reconstruction procedure,
especially because quantities of interest are normally extracted from the
kinematical potential by the application of differential operators on
$\Phi$, a process that is notoriously dangerous when applied to a noisy signal.

\noindent {\bf 3. Our smoothing procedure }
\vspace {0.3 cm}

\noindent {\it 3.1. The philosophy }
\vspace {0.1 cm}

Peculiar radial velocities have been measured for several thousands of
galaxies,
and these have large statistical uncertainties. But the major
difficulty arises from the fact that these galaxies with  observed
radial velocity are not sampled homogeneously throughout space.
The radial velocity field $v_r({\bf x})$ is sampled on a 3-dimensional
spatial support (defined by the positions of galaxies) sparsely
distributed, with large voids empty of information. One thus
has to first smooth the observed radial velocity data
before applying a potential reconstruction procedure. This is
done in the POTENT method [2,4]
by using a tensor window function of large
effective diameter $2500$ $km.
s^{-1}$ compared to the size of the sample (approximatively $10000$ $km.
s^{-1}$). One thus easily understands the importance of the choice
of the smoothing procedure and the attempts
to improve upon this procedure (see [9]).
\\\indent The main weakness we see in the smoothing procedure used
by the POTENT method is that the effective radius of their smoothing
window function is everywhere constant throughout the sample. Fictitious
information is thus added where voids larger than the effective smoothing
radius are present, and information is lost in oversampled regions.
Because of the sparseness of the available data, it seems fundamental to avoid
this loss of information. We develop below
a smoothing scheme based on the use of the wavelet formalism which
allows the radius of the smoothing window function to vary from place
to place in the sample. In terms of wavelet vocabulary (see section 2.1),
we perform the reconstruction of the radial velocity field with
a spatially variable cut-off scale $s_c({\bf x})$, halting the wavelet
reconstruction at a larger scale in undersampled regions and at smaller
scales where the distribution of observed galaxies is dense. Our
aim is to extract a smoothed velocity field containing all the information
present in the catalog of the observed radial velocity of galaxies but no more.
\\\indent The resulting map of the cut-off scale  $s_c({\bf x})$ is wholly
determined by the spatial distribution of the 3-dimensional
support (i.e : the positions $\{{\bf x}_i\}_{i=1,N}$ of the $N$ galaxies of the
catalog). It permits us
to compare our output smoothed radial velocity field
 derived from a known sparsely sampled radial velocity
field $\{v_r({\bf x}_i)\}_{i=1,N}$, with the
wavelet reconstruction of the true field $v_r({\bf x})$ stopped
at the cut-off scale map $s_c({\bf x})$ : $(W_{s_c({\bf x})}v_r)({\bf x})$.
Thus, the errors generated during the successive steps of our smoothing
procedure can be quantified.

\noindent {\it 3.2. The smoothed velocity field }
\vspace {0.1 cm}

The observed radial velocity field we have to smooth is sampled
on a 3-dimensional support inhomogeneously distributed throughout
space. We have simulated a typical cosmic radial velocity field and sampled
it on the support defined by the real positions (expressed in cartesian
supergalactic coordinates) of the galaxies of the MARK II catalog compiled
by D. Burstein (416 independent objects are present). Figure 2 shows the
simulated radial velocity field on nine cuts passing through
a cube of size $10000$ $km.s^{-1}$ and centered on our galaxy (herein
distances
are expressed in $km.s^{-1}$; for a Hubble constant $H_0$ of $100$
$km.s^{-1}/Mpc$, the size of the sample is $100$ $Mpc$). This
non-uniform spatial sampling prevents us in practice from estimating
integrals over space such as those of equation (2.2)
by their associated discrete riemannian
sums over the data positions. However, if we first restore the
uniformity to the spatial support, such approximations can
afterwards be performed with no prejudice, as long as the spatial
extension $s$ of the kernel involved in the convolution of equation (2.2)
is greater than the elementary distance between
neighbouring data points on the homogeneous support. Our smoothing scheme
explores just this possibility.
\begin{figure}
\vspace{9 cm}
\caption{
The simulated radial velocity field $v_r({\bf x}_i)$
at the real positions of the MARK II catalog (the length of arrows is
proportional to the amplitude of the field).
}
\end{figure}
\begin{figure}
\vspace{9 cm}
\caption{
The associated field
$v_r'({\bf \mu}({\bf x}_i))$ in the fictitious $E_{\bf \mu}$ space.
}
\end{figure}
\\\indent We call $E_{\bf x}$ the real 3-dimensional
space wherein the spatial support
$\{{\bf x}_i\}_{i=1,N}$ of the $N$ galaxies of the catalog
is inhomogeneously distributed and we
define by $\rho({\bf x})$ the spatial distribution of the support in
this space. We introduce a mapping ${\bf \mu}$ from this real space
$E_{\bf x}$ into a fictitious 3-dimensional space $E_{\bf \mu}$ such
that the image
$\{{\bf \mu}_i={\bf \mu}({\bf x}_i)\}_{i=1,N}$ of the support under the mapping
${\bf \mu}$ is uniformly distributed in the space $E_{\bf \mu}$ :
$${\bf \mu}\,\,:\,\,\cases{\,E_{\bf x}\,\,
\longrightarrow \,\,E_{\bf \mu}
\cr
\,\, \cr
             \,  {\bf x} \,\, \longmapsto \,\,{\bf \mu}({\bf x}) \cr}
\,\,\,\,\,\,\,\,
J_{\bf \mu}({\bf x})\,\,=\,\,\left|\,det\,\left[ {{\partial \mu_j}\over
{\partial x_k}}({\bf x})
\right]\,\right|\,\,=\,\,\rho({\bf x})
\eqno(3.1)$$
In practice, we evaluate the mapping ${\bf \mu}$ using an algorithm
(this algorithm is described in the appendix A).
\\\indent The first step of our smoothing scheme is to associate to the
set of data $\{v_r({\bf x}_i)\}_{i=1,N}$ of the real space
$E_{\bf x}$, the set $\{v_r'({\bf \mu}_i)\,
=\,v_r'({\bf \mu}({\bf x}_i))\}_{i=1,N}$ in the fictitious
$E_{\bf \mu}$ space such that :
$$v_r'({\bf \mu}({\bf x}))\,\,=\,\,
v_r({\bf x}) \eqno(3.2)$$
This operation is illustrated in figure 3 where
the $\{v_r'({\bf \mu}_i)
\}_{i=1,N}$ are shown inside the normalized cube in the fictitious
$E_{\bf \mu}$ space.
\\\indent We remark that the function $v_r'$ is now sampled on
a uniform support
$\{{\bf \mu}_i={\bf \mu}({\bf x_i})\}_{i=1,N}$ in $E_{\bf \mu}$.
It is therefore possible to define
an elementary cut-off scale $s_\mu$, or minimal resolution length,
as the mean distance between 2
neighbouring points in $E_{\bf \mu}$ , (herein $s_\mu\,=\,0.15$)
and to perform in the $E_{\bf \mu}$
space the wavelet reconstruction $W_{s_\mu}v_r'$ of $v_r'$
halted at the cut-off scale $s_\mu$ (see equation 2.4). The result
of this operation is presented in figure 4.
Each components $v_r'^{(s)}({\bf \mu})$ involved in the integral
over the scales of equation 2.4 is estimated by replacing the spatial
convolution of equation 2.2 by its associated discrete riemannian sum
over the uniformly distributed set of points $\{{\bf \mu}_i\}_{i=1,N}$.
Since the
spatial extension $s$ of the kernel $K$ involved in the convolution is
indeed always greater than the separation between the ${\mu_i}$'s,
this standard estimation operation can be done without creating fictitious
information.
Note that
$W_{s_\mu}v_r'({\bf \mu})$ is now
defined for every ${\bf \mu}$ of $E_{\bf \mu}$ and contains
all of the information that can be extracted from the $\{v_r'({\bf \mu}_i)
\}_{i=1,N}$ data in the fictitious $E_{\bf \mu}$ space.
\\\indent The last step of our smoothing procedure is to return
to the real space $E_{\bf x}$ through the inverse mapping
${\bf \mu}^{-1}$. Our output smoothed velocity field
$(Mv_r)({\bf x})$ is finally the field corresponding to
$W_{s_\mu}v_r'$ in the real space $E_{\bf x}$ namely :
$$(Mv_r)({\bf x})\,\,=\,\,(W_{s_\mu}v_r')({\bf \mu}({\bf x})) \eqno(3.3)$$
This operation is illustrated figure 5. Since the mapping
$\mu$ establishes a one-to-one correspondance between points of the real space
$E_{\bf x}$ and those of the fictitious  $E_{\bf \mu}$, our output smoothed
radial velocity field
$(Mv_r)({\bf x})$ contains the maximum of information which can
be extracted from the $\{v_r'({\bf \mu}_i)
\}_{i=1,N}$ and thus from the observed radial velocities
$\{v_r({\bf x}_i)
\}_{i=1,N}$ of the $N$ galaxies in the catalogue. In this way
our smoothing procedure is minimal (no loss of information).
\begin{figure}
\vspace{9 cm}
\caption{
The wavelet reconstruction
$(W_{s_\mu}v_r')(\bf \mu)$ of $v_r'$ halted at the cut-off
scale $s_\mu$.
}
\end{figure}
\begin{figure}
\vspace{9 cm}
\caption{
Our output smoothed radial velocity field
$(Mv_r)(\bf x)$ (in the real space $E_{\bf x}$).
}
\end{figure}

\noindent {\it 3.3. Correspondence with a theoretical quantity }
\vspace {0.1 cm}

Thanks to the scale resolution  properties of the wavelet transform,
our output smoothed radial velocity field $(Mv_r)({\bf x})$
can be expressed as a theoretical quantity.
We prove in appendix B that, so long as the mapping ${\bf \mu}$ satisfies
a validity condition (see below equation 3.5), the following equality holds :
$$(Mv_r)({\bf x})\,\,=\,\,(W_{s_c({\bf x})}v_r)({\bf x})
\,\,\,\,\,\,{\rm with}\,\,\,\,\,\,s_c({\bf x})\,\,=
\,\,{{s_{\mu}}\over{\rho({\bf x})^{1/3}}} \eqno(3.4)$$
Thus our output smoothed radial velocity field may be identified with
the wavelet reconstruction of $v_r$ halted at the inverse map of the
cut-off scale
$s_c({\bf x})$ which varies with ${\bf x}$. We show in figure 6 the cut-off
scale  inverse map corresponding to the spatial distribution previously
presented
in figure 2. The value of $s_c({\bf x})$ is derived from the jacobian
associated with the mapping ${\bf \mu}$ (see appendix A).
The lower the density at the position ${\bf x}$, the larger is the
corresponding cut-off scale.
\\\indent
We exhibit in figure 7 the wavelet reconstruction of the previously simulated
radial velocity field halted at the cut-off scale  $s_c({\bf x})$.
We notice that even if the main features remain, our smoothed radial velocity
field $(Mv_r)({\bf x})$ of figure 5 differs in detail
from $(W_{s_c({\bf x})}v_r)({\bf x})$. The reason
for this discrepancy is that the mapping ${\bf \mu}$ doesn't in fact satisfy
the validity condition (appendix B) which stipulates that for every ${\bf x}$
and
vector ${\bf h}$ :
$${\rm if}\,\,\,\|{\bf h}\|\,\le\,s_c({\bf x})\,,\,\,\,\,\,\,\,\,\,\,
\,\,\,\,\left\|\,\left[ {{\partial \mu_j}\over
{\partial x_k}}({\bf x})
\right].[{\bf h}]\,\right\|\,\,\approx\,\,
\left|\,det\,\left[ {{\partial \mu_j}\over
{\partial x_k}}({\bf x})
\right]\,\right|^{1/3} \times \|{\bf h}\| \eqno(3.5)$$
or in other words that the mapping ${\bf \mu}$ is locally equivalent
to a rotation-dilation transformation (see appendix B).
\begin{figure}
\vspace{9 cm}
\caption{
Isocontours of the cut-off scale map
$s_c(\bf x)$ (Normal, heavy and dotted contours are resp.,
$750$, $1500$ and $2250$ $km.s^{-1}$).
}
\end{figure}
\begin{figure}
\vspace{9 cm}
\caption{
The wavelet reconstruction
$(W_{s_c({\bf x})}v_r)(\bf x)$ of the simulated radial velocity field
$v_r({\bf x})$ halted at the cut-off
scale map $s_c({\bf x})$.
}
\end{figure}
\\\indent However, this discrepancy between our output smoothed
velocity field $(Mv_r)({\bf x})$ and the wavelet reconstruction
$(W_{s_c({\bf x})}v_r)({\bf x})$
of the true radial velocity field $v_r({\bf x})$ halted at the
cut-off scale map $s_c({\bf x})$ can be quantified by analysing
the spatial distribution of the $N$ galaxies of the sample. Thus at
each position ${\bf x}$ in the sampled volume, we can define
a "relative sampling error" $E({\bf x})$ computed as follows :
$$E^2({\bf x})\,\,=\,\,{{\sum_{i=1}^N\,\left[K(s_\mu,{\bf \mu}({\bf x}),
{\bf \mu}({\bf y}_i))\,-\,1/\rho({\bf y}_i)\,K(s_c({\bf x}),
{\bf x},
{\bf y}_i)\right]^2}\over{
\sum_{i=1}^N\,\left[K(s_\mu,{\bf \mu}({\bf x}),
{\bf \mu}({\bf y}_i))
\right]^2}} \eqno(3.6)$$
This `relative sampling error map' $E({\bf x})$ gives at each
position ${\bf x}$ the relative error
between $(Mv_r)({\bf x})$ and
$(W_{s_c({\bf x})}v_r)({\bf x})$
as a fraction of the amplitude of the true radial velocity field $v_r({\bf
x})$.
The relative sampling error map $E({\bf x})$ for the spatial distribution
of the galaxies of the MARK II catalogue is shown in figure 8.
In practice we can discard the regions of the sample where the sampling
\begin{figure}
\vspace{9 cm}
\caption{
Isocontours of the
relative sampling errors map $E({\bf x})$
(Normal, heavy and dotted contours are resp.,
$15\%$, $25\%$ and $50\%$).
}
\end{figure}
errors are too high.

\noindent {\bf 4. The kinematical potential} ${\Phi}({\bf x})\,\,=
\,\,(Pv_r)({\bf x})$
\vspace {0.3 cm}

We have shown in section 3.3 that, in the regions of
the sampled volume where the validity
condition is verified (i.e : the domains where the relative sampling errors
$E({\bf x})$ are low), our smoothed radial velocity field
$(Mv_r)({\bf x})$ can be identified with
$(W_{s_c({\bf x})}v_r)({\bf x})$. As we suggested in the introduction,
there are good reasons to believe that the cosmic velocity field
is irrotational up to an ill-defined scale and thus that a kinematic
potential $\Phi({\bf x})$
can be inferred by integrating the cosmic radial velocity
field $v_r({\bf x})$ along the line-of-sight (equation 1.1). It
turns out that the ill-defined scale of irrotationality is generally
presumed to be smaller than the cut-off scales $s_c({\bf x})$ which limit
the spatial resolution of the accessible cosmic radial velocity field
$(W_{s_c({\bf x})}v_r)({\bf x})$. It becomes thus possible to extract
the velocity potential $\Phi({\bf x})$ by applying
the `integral-along-the-line-of-sight' operator $P$ either to
$(W_{s_c({\bf x})}v_r)({\bf x})$ or to the smoothed radial velocity
field $(Mv_r)({\bf x})$.
\begin{figure}
\vspace{9 cm}
\caption{
Isocontours of the potential of
the smoothed radial velocity field
$(P\circ W_{s_c({\bf x})}v_r)(\bf x)$ (the contour spacing is $2.5\,
10^5$ $(km.s^{-1})^2$, negative contours are dotted, heavy contour is $0$).
}
\end{figure}
\begin{figure}
\vspace{9 cm}
\caption{
Isocontours of the smoothed simulated
potential
$(W_{s_c({\bf x})}\circ Pv_r)(\bf x)$ (the contour spacing is $2.5\,
10^5$ $(km.s^{-1})^2$, negative contours are dotted, heavy contour is $0$).
}
\end{figure}
\\\indent Unfortunately,
this operator $P$ has a non-local
character. Hence, the potential derived from the smoothed radial
velocity field $P\circ W_{s_c({\bf x})}v_r$ differs from the smoothed
potential of the velocity field $W_{s_c({\bf x})}\circ Pv_r$ (and of
course from the non-smoothed kinematical potential
${\Phi}\,\,= \,\,Pv_r$)
(see appendix C). We illustrate this discrepancy by plotting in figure 9
the potential
derived from $(W_{s_c({\bf x})}v_r)({\bf x})$ and in figure 10 the smoothed
simulated potential $(W_{s_c({\bf x})}{\Phi})({\bf x})$.
We want to emphasize that this behaviour is not due to the way
we smooth the radial velocity field but is intrinsically linked to the
nature of the operator $P$.
\\\indent The point has its importance since
the kinematical potential (or the reconstructed 3-dimensional velocity
field) derived from catalogues of radial peculiar velocities is
often considered as reference data for other studies (for example
Saunders et {\it al.}
(1991), Dekel et {\it al.} (1993)). As an example note that,
above the scale of irrotationality,
the total mass density perturbation field $\delta_t({\bf x})$ is linked
to the kinematical potential $\Phi({\bf x})$ through the Poisson
equation ($\delta_t({\bf x})\,\,\propto\,\,\nabla^2\Phi({\bf x})$).
Adopting the hypothesis that the luminous matter traces the mass
($\delta_t({\bf x})\,\propto\,\delta_l({\bf x})$),
a smoothed total mass density field may be extracted from observations of the
spatial distribution of galaxies. Unfortunately for the intercomparison of such
 independent results :
$$(W_{s_c}\delta_t)({\bf x})\,\,=\,\,(\nabla^2(W_{s_c}\circ Pv_r))({\bf x})
\,\,{\bf \ne}\,\,(\nabla^2(P\circ W_{s_c}v_r))({\bf x}) \eqno(4.1)$$
\\\indent One thus should be very cautious
when comparing the density or kinematical potential
derived from observed radial peculiar velocity catalogues with similar
quantities
obtained from other studies, such as those based on number galaxy counts.

\noindent {\bf 5. Conclusions}
\vspace {0.3 cm}

We have presented a method, based on the properties of the wavelet transforms,
for smoothing a field sampled on a support inhomogeneously distributed
throughout the space of interest. Our smoothing scheme is minimal (no loss of
information) and our output smoothed field can be identified
with a well-defined theoretical quantity, as long as the spatial support
of the field satisfies certain criteria. We expect this technique to be
of quite general importance. The particular application of this smoothing
scheme to the observed cosmic radial velocity field discussed above, reveals
some apparently previously unrecognized
limitations concerning the reconstruction of the kinematic potential
from the smoothed radial velocity field. Indeed, we prove that
this potential will generally differ a priori
from a similar quantity
obtained from other cosmological studies. The existence of this kind
of error now has to be taken into account.

\noindent {\bf Acknowledgments}
\vspace {0.1 cm}

S. Rauzy wants to recognize the hospitality of the Physics department of
Queen's University, Kingston, Ontario, where a large part of this
work was achieved.

\noindent {\bf Appendix A : The $\bf \mu$ mapping }
\vspace {0.1 cm}

In practice, we evaluate the mapping ${\bf \mu}$ using
the following algorithm.
We first choose a direction (the $x_1$-axis for example). Along this
direction, we sort the $N$ points of the sample by the increasing
value of their $x_1$-coordinate. To the object of rank $\it i$, we
then affect the value $\mu_1\,=\,i/N$ to its $\mu_1$-coordinate, the
first component of
the mapping ${\bf \mu}$. This procedure ensures us that the set of the
$\mu_1$-coordinates
of the sampled objects are uniformly distributed along
the $\mu_1$-axis in the fictitious space $E_{\bf \mu}$. We now
divide the cube in slices perpendicular to the $x_1$-axis (7 $x_1$-slices
for this sample) such that
each slice contains approximatively the same number $N_1$ of data points.
The non-uniform sampling in the $E_{\bf x}$ space
implies that each slice have its own width. Nevertheless, the
widths of these slices become equal along the $\mu_1$-axis in the space
$E_{\bf \mu}$.
\\\indent The next step of our algorithm is to repeat this construction for
each $x_1$-slice, choosing the $x_2$-axis as the sorting direction.
Inside each $x_1$-slice, the object of rank $\it j$ has a $\mu_2\,=\,
j/N_1$. We thus
obtain $\mu_2$-coordinates
uniformly distributed inside each $x_1$-slice, and so for the set
of all slices
in the fictitious $E_{\bf \mu}$ space.
The $x_1$-slices are afterwards
divided again in slices perpendicular to the $x_2$-axis (7 for this sample)
containing the same
number $N_2$ of objects and each $x_1x_2$-slice is sorted along
the $x_3$-axis (for an object of rank $\it k$, $\mu_3\,=\,k/N_2$).
$N_2$ is 8 or 9 in our case, depending on the $x_1x_2$-slice treated.
\\\indent At the output of this algorithm, we then obtain a set
of $N$ points $\{{\bf \mu}_i\,=\,{\bf \mu}({\bf x_i})\}_{i=1,N}$
uniformly distributed inside a normalized cube in the fictitious
space $E_{\bf \mu}$. A continuous version of the ${\bf \mu}$ mapping
as well as its inverse mapping ${\bf \mu}^{-1}$
can be computed, if necessary, by using a linear interpolation
on the $\mu_1$, $\mu_2$ and $\mu_3$ components of the transformation
${\bf \mu}$.

In order to evaluate quantities such as the cut-off scale map
$s_c({\bf x})$ (section 3.3.), the spatial density distribution of the support
$\rho({\bf x})$ in the real space $E_{\bf x}$ has to be
evaluated. We then first estimate at the position ${\bf x}$ the 9 terms
$({\partial \mu_j}/
{\partial x_k})({\bf x})$ ($j$ and $k$ vary from 1 to 3) of the
transformation coordinate matrix associated with the mapping ${\bf \mu}$.
We then derive from this matrix its jacobian $J_{\bf \mu}({\bf x})$
and finally adopt for the density distribution $\rho({\bf x})$
a smoothed version of $J_{\bf \mu}({\bf x})$.

\noindent {\bf Appendix B : Comparison between
$(Mv_r)({\bf x})$ and $(W_{s_c({\bf x})}v_r)({\bf x})$}
\vspace {0.1 cm}

Our goal here is to identify our output smoothed radial velocity field
$(Mv_r)({\bf x})$ with a theoretical quantity derived from the true radial
velocity field $v_r({\bf x})$. Our smoothing procedure ensures
us that the following equality holds (see equation 3.3) :
$$(Mv_r)({\bf x})\,\,=\,\,(W_{s_\mu}v_r')({\bf \mu}({\bf x})) \eqno(B.1)$$
In the fictitious $E_{\bf \mu}$ space,
$(W_{s_\mu}v_r')({\bf \mu})$ is the wavelet reconstruction
of the field $v_r({\bf \mu})$ halted at the cut-off scale $s_\mu$.
It satisfies (see equations 2.3, 2.4 and 2.5) :
$$v_r'({\bf \mu})\,\,=\,\,Wv_r'({\bf \mu})\,\,=\,\,W_{s_\mu}v_r'({\bf \mu})\,\,
+\,\,W^{s_\mu}v_r'({\bf \mu}) \eqno(B.2)$$
Now, Let us focus on the measurable component
$(W^{s_\mu}v_r')({\bf \mu}({\bf x}))$. In the  fictitious $E_{\bf \mu}$
space, it can be expressed as follows (see equations 2.2 and 2.5) :
$$(W^{s_\mu}v_r')({\bf \mu}({\bf x}))
\,\,=\,\,\int_0^{s_\mu}\,{{ds}\over{s}}\,\,
\int_{E_{\bf \mu}}\,d^3{\bf \tau}
\,v_r'({\bf \tau})\,K(s,{\bf \mu}({\bf x}),{\bf \tau}) \eqno(B.3)$$
Recalling equation 3.2 and  introducing ${\bf y}$ such that
${\bf \tau}\,=\,{\bf \mu}({\bf y})$, this equation becomes in
the real space $E_{\bf x}$ :
$$(W^{s_\mu}v_r')({\bf \mu}({\bf x}))
\,\,=\,\,\int_0^{s_\mu}\,{{ds}\over{s}}\,\,
\int_{E_{\bf x}}\,\rho({\bf y})\,d^3{\bf y}
\,v_r({\bf y})\,K(s,{\bf \mu}({\bf x}),{\bf \mu}({\bf y})) \eqno(B.4)$$
At this stage, we have to use the specific properties of
the reproducing kernel $K$.
This kernel $K$ can be derived from an "analysing"
function $k$ of unit spatial extension due to
the following symetry (see [7]) :
$$ K(s,{\bf x},{\bf y})\,\,=\,\,{1\over{s^3}}\,k\left({{\|{\bf x}-{\bf y}
\|}\over{s}}\right) \eqno(B.5)$$
Hence, for every real number $\alpha$, the kernel
satisfies :
$$ K(s,\alpha{\bf x},\alpha{\bf y})\,\,=\,\,
{1\over{\alpha^3}}\,
K(s/\alpha,{\bf x},{\bf y}) \eqno(B.6)$$
Keeping this in mind, we develop the mapping ${\bf \mu}$ to the
first order of its vectorial taylor series with respect to
${\bf x}$ or ${\bf y}$. The difference ${\bf \mu}({\bf x})-
{\bf \mu}({\bf y})$ involved in the kernel $K$ intervening
in equation B.4 can then be approximated :
$$\|{\bf \mu}({\bf x})\,-\,{\bf \mu}({\bf y})\|\,\,\approx
\,\,\left\|\,\left[ {{\partial \mu_j}\over
{\partial x_k}}({\bf x})
\right].[{\bf x}-{\bf y}]\,\right\|\,\,\approx
\,\,\left\|\,\left[ {{\partial \mu_j}\over
{\partial x_k}}({\bf y})
\right].[{\bf x}-{\bf y}]\,\right\| \eqno(B.7)$$
This approximation is in practice performed inside regions of
the $E_{\bf \mu}$ space not larger than the cut-off scale $s_\mu$
(these regions are in fact
defined by the spatial extension of the kernel $K$ involved in
the spatial convolution of the equation B.4).
\\\indent The next step is to make the crucial assumption that the
mapping $\mu$ can be locally identified with a rotation-dilation
transformation, or in other words that
for $\|{\bf h}\|\,\le\,s_\mu/\rho^{1/3}({\bf x})$ :
$$\left\|\,\left[ {{\partial \mu_j}\over
{\partial x_k}}({\bf x})
\right].[{\bf h}]\,\right\|\,\,\approx\,\,
\left|\,det\,\left[ {{\partial \mu_j}\over
{\partial x_k}}({\bf x})
\right]\,\right|^{1/3} \times \|{\bf h}\|\,\,=\,\,
\rho({\bf x})^{1/3} \times \|{\bf h}\| \eqno(B.8)$$
This equation is certainly not valid for every point ${\bf x}$
of the sampled volume. Nevertheless, in the regions of space where
this validity condition does hold, the following approximation is
valid (see equations B.5, B.6, B.7 and B.8) :
$${\rm if}\,\,\,s\,\le\,s_\mu,
\,\,\,\,\,\,\,\,\,\,\,\,\,\,\,\,\,\,\,
K(s,{\bf \mu}({\bf x}),{\bf \mu}({\bf y}))\,\,\approx\,\,
{1\over{\rho({\bf y})}}\,
K(s/\rho({\bf x})^{1/3},{\bf x},{\bf y}) \eqno(B.9)$$
Then we can define a cut-off scale map $s_c({\bf x})$ in the $E_{\bf x}$
space :
$$ s_c({\bf x})\,\,=
\,\,{{s_{\mu}}\over{\rho({\bf x})^{1/3}}} \eqno(B.10)$$
and it turns out that if we introduce $S$ such that
$S\,=\,s/\rho({\bf x})^{1/3}$, equations B.4, B.9 and B.10 give :
$$(W^{s_\mu}v_r')({\bf \mu}({\bf x}))
\,\,\approx\,\,\int_0^{s_c({\bf x})}\,{{dS}\over{S}}\,\,
\int_{E_{\bf x}}\,d^3{\bf y}
\,v_r({\bf y})\,K(S,{\bf x},{\bf y})
\,\,=\,\,(W^{s_c({\bf x})}v_r)({\bf x})
\eqno(B.11)$$
Finally, because of equations 2.3, B.1, B.2 and B.11, the following equality
holds in the regions of the sampled volume where the validity condition
of equation B.8 is satisfied :
$$(Mv_r)({\bf x})\,\,=\,\,(W_{s_c({\bf x})}v_r)({\bf x}) \eqno(B.12)$$
\\\indent Thus, our output smoothed radial velocity field
$(Mv_r)({\bf x})$ can be compared to the wavelet reconstruction
$(W_{s_c({\bf x})}v_r)({\bf x})$ of the real radial velocity field
$v_r({\bf x})$ halted at the cut-off scale map $s_c({\bf x})$.
In practice we evaluate the "degree of approximation" of the validity
condition by comparing the 2 sides of the equation B.9 integrated over
the space at the fixed scale $s_\mu$. This quantity $E({\bf x})$ is
defined as follows :
$$E^2({\bf x})\,\,=\,\,{{\sum_{i=1}^N\,\left[K(s_\mu,{\bf \mu}({\bf x}),
{\bf \mu}({\bf y}_i))\,-\,1/\rho({\bf y}_i)\,K(s_c({\bf x}),
{\bf x},
{\bf y}_i)\right]^2}\over{
\sum_{i=1}^N\,\left[K(s_\mu,{\bf \mu}({\bf x}),
{\bf \mu}({\bf y}_i))
\right]^2}} \eqno(B.13)$$
This relative sampling errors map $E({\bf x})$ gives at each
position ${\bf x}$ the relative error
between $(Mv_r)({\bf x})$ and
$(W_{s_c({\bf x})}v_r)({\bf x})$
with respect to the amplitude of the true radial velocity field $v_r({\bf x})$.

\noindent {\bf Appendix C : $P\circ W_{s_c({\bf x})}v_r\,\,\ne\,\,
W_{s_c({\bf x})}\circ Pv_r$}
\vspace {0.1 cm}

We prove in this appendix that the `integral-along-the-line-of-sight
operator' $P$ doesn't commute with the necessary smoothing operation
that must first be performed on the radial velocity field $v_r({\bf x})$.
This behaviour doesn't depend on the particular nature of our preliminary
smoothing, rather it remains true more generally even if for example
 the effective radius
of the smoothing window function is constant throughout the sampled
volume as in POTENT.
\\\indent Let us express the potential derived from the smoothed radial
velocity field $P\circ W_{s_c}v_r$ in terms of the smoothing
window function introduced equation 2.6 (herein , $s_c$ is chosen
constant throughout the space since otherwise our argument is even stronger) :
$$(P\circ W_{s_c}v_r)({\bf x})
\,\,=\,\,\int_0^{1}\,dl\,\,\,\,
\int_{E_{\bf x}}\,d^3{\bf y}
\,\,v_r({\bf y})\,g(s_c,l{\bf x},{\bf y})
\eqno(C.1)$$
$(P\circ W_{s_c}v_r)({\bf x})$ is the integral along the line-of-sight of the
field $v_r$ smoothed with a window function of constant effective
radius $s_c$ from ${\bf O}$ to ${\bf x}$.
\\\indent In the same way, let us express the smoothed version
at the scale $s_c$ of the potential
$(W_{s_c}\circ Pv_r)({\bf x})$ :
$$(W_{s_c}\circ Pv_r)({\bf x})
\,\,=\,\,\int_{E_{\bf x}}\,d^3{\bf y}
\,\,\,\,\int_0^{1}\,dl\,
v_r(l{\bf y})\,\,\,g(s_c,{\bf x},{\bf y})
\eqno(C.2)$$
Because of the specific properties of the wavelet smoothing window function,
it turns out that (see equation 2.6, B.5 and B.6) :
$$ g(s_c,{\bf x},{\bf y})\,\,=\,\,
l^3\,g(ls_c,l{\bf x},l{\bf y}) \eqno(C.3)$$
Hence equation C.2 can be re-expressed, if we introduce ${\bf Y}$ such
that ${\bf Y}\,=\,l{\bf y}$ :
$$(W_{s_c}\circ Pv_r)({\bf x})
\,\,=\,\,\int_0^{1}\,dl\,\,\,\,
\int_{E_{\bf x}}\,d^3{\bf Y}
\,\,v_r({\bf Y})\,g(ls_c,l{\bf x},{\bf Y})
\eqno(C.4)$$
$(W_{s_c}\circ Pv_r)({\bf x})$
is then the integral along the line-of-sight of the
field $v_r$ smoothed with a window function with a variable effective
radius $ls_c$ all along the line-of-sight, from 0 at the observator
position ${\bf O}$ to $s_c$ in ${\bf x}$.
It thus follows that :
$$P\circ W_{s_c}v_r\,\,\ne\,\,
W_{s_c}\circ Pv_r \eqno(C.5)$$
as was required.

\vspace {0.2 cm}
\noindent {\bf References }
\vspace {0.1 cm}

\noindent [1] Bertschinger E. and Dekel A. (1989) {\it Astrophys. J.
(Letters)}, {\bf 336}, L5.
\\\noindent [2] Bertschinger E., Dekel A., Faber S.M. Dressler A. and
Burstein D. (1990) {\it Astrophys. J.}, {\bf 364}, 370.
\\\noindent [3] Daubechies I. (1992) in {\it Ten lectures on Wavelets},
CBMS-NSF Regional Conference Series in Applied Mathematics, SIAM.
\\\noindent [4] Dekel A., Bertschinger E. and Faber S.M.
(1990) {\it Astrophys. J.}, {\bf 364}, 349.
\\\noindent [5] Dekel A., Bertschinger E., Yahil A., Strauss M., Davis M.
and Huchra J.
(1993) {\it Astrophys. J.}, {\bf 412}, 1.
\\\noindent [6] Grossmann A., Kronland-Martinet R. and Morlet J. (1990), in
{\it Wavelets}, Proc. of the International Conference, Tchamitchian
ed. , Springer-Verlag.
\\\noindent [7] Rauzy S., Lachi\`eze-Rey M. and Henriksen R.N. (1993)
{\it Astr. Astrophys.}, {\bf 273}, 357.
\\\noindent [8] Saunders W., Rowan-Robinson M., Lawrence A.,
Crawford J., Ellis R., Frenk C.S., Parry I., Xiaoyang X.,
Allington-Smith J., Estathiou G. and Kaiser N. (1991) {\it Nature},
{\bf 349}, 32.
\\\noindent [9] Simmons J.F.L., Newsam A. and Hendry M.A. (1994)
{\it Astr. Astrophys.}, in press.

\end{document}